\documentclass[prl,10pt,notitlepage,twocolumn,superscriptaddress]{revtex4}

\usepackage{graphicx}
\usepackage{amsmath}
\usepackage{amssymb}
\usepackage{amsfonts}
\usepackage{color}
\usepackage{xspace}
\usepackage{subfigure}
\usepackage{bm}
\usepackage{bbold}
\usepackage{blkarray}
\usepackage{pdflscape}

\newcommand{\bra}[1]{\langle #1 |}
\newcommand{\ket}[1]{| #1 \rangle}

\newcommand{\etal}{\emph{et al}.\xspace}
\newcommand{\Cllong}{$\kappa$-(BEDT-TTF)$_2$Cu[N(CN)$_2$]Cl\xspace}
\newcommand{\Cl}{$\kappa$-Cl\xspace}
\newcommand{\CNlong}{$\kappa$-(BEDT-TTF)$_2$Cu$_2$(CN)$_3$\xspace}
\newcommand{\Aglong}{$\kappa$-(BEDT-TTF)$_2$Ag$_2$(CN)$_3$\xspace}
\newcommand{\Sbonelong}{$\beta$'-EtMe$_3$Sb[Pd(dmit)$_2$]$_2$\xspace}
\newcommand{\CN}{CuCN\xspace}
\newcommand{\Ag}{AgCN\xspace}
\newcommand{\kX}{$\kappa$-(BEDT-TTF)$_2$\textit{X}\xspace}
\newcommand{\DM}{Dzyaloshinskii-Moriya\xspace}

\begin{document}

\title{Dynamical  reduction of the dimensionality of exchange interactions and the ``spin-liquid'' phase of $\kappa$-(BEDT-TTF)$_2X$} 

\author{B. J. Powell}
\email{bjpowell@gmail.com}
\affiliation{School of Mathematics and Physics, The University of Queensland, Brisbane, Queensland, 4072, Australia}

\author{E. P. Kenny}\affiliation{School of Mathematics and Physics, The University of Queensland, Brisbane, Queensland, 4072, Australia}
%
%

\author{J. Merino}
\affiliation{Departamento de F\'isica Te\'orica de la Materia Condensada, Condensed Matter Physics Center (IFIMAC) and
	Instituto Nicol\'as Cabrera, Universidad Aut\'onoma de Madrid, Madrid 28049, Spain}

\begin{abstract}
We show that the anisotropy of the effective spin model for the dimer Mott insulator phase of  $\kappa$-(BEDT-TTF)$_2X$ salts is dramatically different from that of the underlying tight-binding model. Intra-dimer quantum interference  results in a  model of coupled spin chains, where frustrated interchain interactions suppress long-range magnetic order. Thus, we argue, the ``spin liquid'' phase observed in some of these materials is a remnant of the Tomonaga-Luttinger physics of a single chain. This is consistent with previous experiments and  resolves some outstanding puzzles.
An erratum \cite{erratum} is added as an appendix.
\end{abstract}

\maketitle

Layered organic charge transfer salts show a wide range of exotic physics due to  strong electronic correlations and geometrical frustration \cite{RPP}. This includes unconventional superconductivity, incoherent metallic transport, multiferroicity, and antiferromagnetism. However, the putative spin liquid states in \CNlong \cite{Shimizu}, \Aglong \cite{Shimizu16} (henceforth, \CN and \Ag respectively) and \Sbonelong \cite{Itou} are, perhaps, the least  understood of these. 

\CN is usually discussed in terms of the nearly triangular Heisenberg model 
\cite{RPP,Balents}.  Here we demonstrate that the theoretical arguments that lead to this model are fallacious. They fail to account for quantum interference within the (BEDT-TTF)$_2$ dimer.  We derive the correct low-energy model including these effects and show that it leads to an anisotropic triangular lattice in the quasi-one-dimensional (q1D) regime, $J_1>J_2$, Fig. 1c. Thus, the spin model for the Mott dimer insulating phases of the organic charge transfer salts are remarkably similar to that describing Cs$_2$CuBr$_4$ and Cs$_2$CuCl$_4$ \cite{Radu-model}, where deconfined spinons have been observed \cite{Radu,Balents}. Our results provide natural explanations for several previously puzzling experiments on the organics.

\begin{figure}
    \begin{minipage}[t]{0.6\columnwidth}
        {\includegraphics[width=\textwidth]{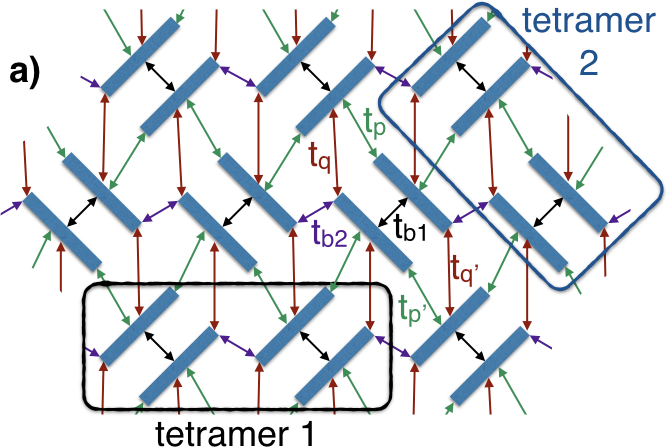}}
    \end{minipage}%
    \begin{minipage}[b]{0.3\columnwidth}
        {\includegraphics[width=\textwidth]{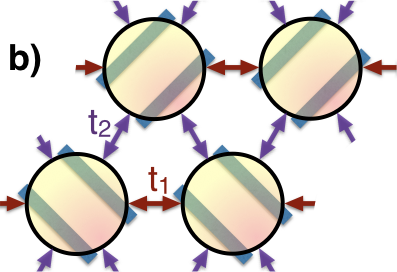}}
        \vfill
        {\includegraphics[width=\textwidth]{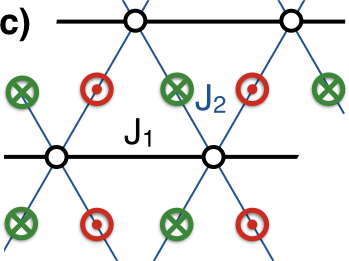}}
    \end{minipage}%
\caption{Models of organic charge transfer salts: (a) Hopping integrals between monomers (bars). To an excellent approximation $t_q=t_{q'}$ and $t_p=t_{p'}$ \cite{Koretsune}. (b) The dimer model. (c)  Heisenberg model in the  dimer Mott insulator phase, the staggered interlayer component of DM interaction is also indicated -- we adopt the convention that the leftmost spin appears first in the DM interaction, ${\bm D}_{ij}\cdot{\bm S}_i\times{\bm S}_j$. } 
\label{fig:models}
\end{figure}

Electronic structure calculations demonstrate that a single molecular orbital contributes to the low-energy process in the \kX salts \cite{RPP,Kino,RossReview,Koretsune}, and that the band structure is described by the tight-binding `monomer model' sketched in Fig. \ref{fig:models}a at three quarters filling. This model is dimerised: $t_{b1}\gg t_{b2}, t_p, t_q$. 
At ambient  pressure \CN, \Ag and \Cllong (henceforth \Cl) display a Mott dimer insulating phase, where excitations away from exactly one hole per dimer are bound \cite{RPP}.

Electronic correlations arise from the  Coulombic repulsion between two holes on the same monomer, $U_m$, or dimer, $V_m$. Thus, the effective Hamiltonian for the $i$th dimer is
$ {\cal H}_{b1}^{(i)} = -t_{b1}\sum_{\sigma} (\hat c_{i1\sigma}^\dagger  \hat c_{i2\sigma} + H.c.) 
+ U_m\sum_\mu \hat n_{i\mu\uparrow}  \hat n_{i \mu\downarrow}
+V_m \hat n_{i 1}  \hat n_{i 2} $,
where  $\hat c_{i\mu\sigma}^{(\dagger)}$ annihilates (creates) an electron with spin $\sigma$ on the $\mu$th monomer of the $i$th dimer, $\hat n_{i\mu\sigma}=\hat c_{i\mu\sigma}^{\dagger} \hat c_{i\mu\sigma}$, and $\hat n_{i\mu}=\sum_\sigma\hat n_{i\mu\sigma}$. Other  Coulomb matrix elements can also be included, but do not  qualitatively change our results and  are neglected below. The hopping between dimers is given by
${\cal H}_{1} = -t_{b2}\sum_{\langle i,j\rangle\sigma}(\hat T_{21}+\hat T_{21}^\dagger)$ and
${\cal H}_{2} = \sum_{[ i,j]\sigma} 
[
	-t_p (\hat T_{21}+\hat T_{21}^\dagger) 
	-t_q (\hat T_{22}+\hat T_{22}^\dagger)
]$,
where $\hat T_{\nu\mu} = \hat c_{i\nu\sigma}^\dagger  \hat c_{j\mu\sigma}$,
$\langle i,j\rangle$ implies a pair of dimers  equivalent to tetramer 1 (Fig. \ref{fig:models}a), and $[ i,j]$ implies a pair of dimers such as tetramer 2.

Kino and Fukuyama (KF) showed that for large enough $U_m$ an insulating phase emerges \cite{Kino}. They argued that this could be understood as a dimer Mott insulator: if one integrates out the bonding combination of molecular orbitals this leaves an effective  half-filled model containing only the antibonding combination of molecular orbitals, $a_{i\sigma}^{\dagger}=\frac1{\sqrt{2}}(c_{i1\sigma}^{\dagger}-c_{i2\sigma}^{\dagger})$. The `dimer model' is 
${\cal H}_{d} = -t_{1}\sum_{\langle i,j\rangle\sigma} (\hat a_{i\sigma}^\dagger  \hat a_{j\sigma} + H.c.) -t_{2}\sum_{[ i,j]\sigma} (\hat a_{i\sigma}^\dagger  \hat a_{j\sigma} + H.c.) 
+ U_d\sum_i \hat a_{i\uparrow}^{\dagger} \hat a_{i\uparrow} \hat a_{i\downarrow}^{\dagger} \hat a_{i\downarrow}$,
Fig. \ref{fig:models}b, where, for $V_m=0$,   $t_1/t_{b2}=t_2/(t_p+t_q)=\sqrt2(\cos\theta-\sin\theta)/4$, and $\tan\theta=(U_m/4t_{b1})-\sqrt{1+(U_m/4t_{b1})^2}$  \cite{RossReview}.  KF estimated the effective interaction between two holes on the same dimer as 
$U_d=E_0(0)+E_0(2)-2E_0(1) = 2t_{b1}+({U_m}/{2})[1-\sqrt{1+({4t_{b1}}/{U_m})}]\simeq2t_{b1}$ for $U_m\gg 4t_{b1}$, 
where $E_0(N)$ is the ground state of the dimer with $N$ holes. 
The $V_m\ne0$ case is discussed in \cite{ScrivenPRB}.

In the Mott dimer phase KF's dimer model reduces to a Heisenberg model, Fig. \ref{fig:models}c:
\begin{eqnarray}
{\cal H}_{H} &=& J_{1}\sum_{\langle i,j\rangle\sigma} \hat{\mathbf{S}}_i \cdot \hat{\mathbf{S}}_j +J_{2}\sum_{[ i,j]\sigma} \hat{\mathbf{S}}_i \cdot \hat{\mathbf{S}}_j,
\label{Eq:Heis}
\end{eqnarray} 
where $\hat{\mathbf{S}}_i$ is the spin operator on the $i$th dimer, 
and in the dimer model $J_{1}=4t_1^2/U_d$ and $J_{2}=4t_2^2/U_d$.

Two decades of  research have been based on these ideas. Thus, it is surprising that no one appears to have asked whether the same parameters  for the Heisenberg model, Eq. (\ref{Eq:Heis}), are found from both the  monomer and dimer models. We do. The answer is no. 

To calculate $J_1$ we perform a canonical transformation \cite{Amie,Jaime17,DiracQ} taking ${\mathcal H}_0=\sum_i{\mathcal H}_{b1}^{(i)}$ as our unperturbed Hamiltonian with the perturbation given by ${\mathcal H}_1$. We  retain terms ${{O}(t_{b2}^2)}$ yielding the interaction described by the first term in Eq. (\ref{Eq:Heis}).  

\begin{figure}
	\begin{center}
		\includegraphics[width=0.8\columnwidth]{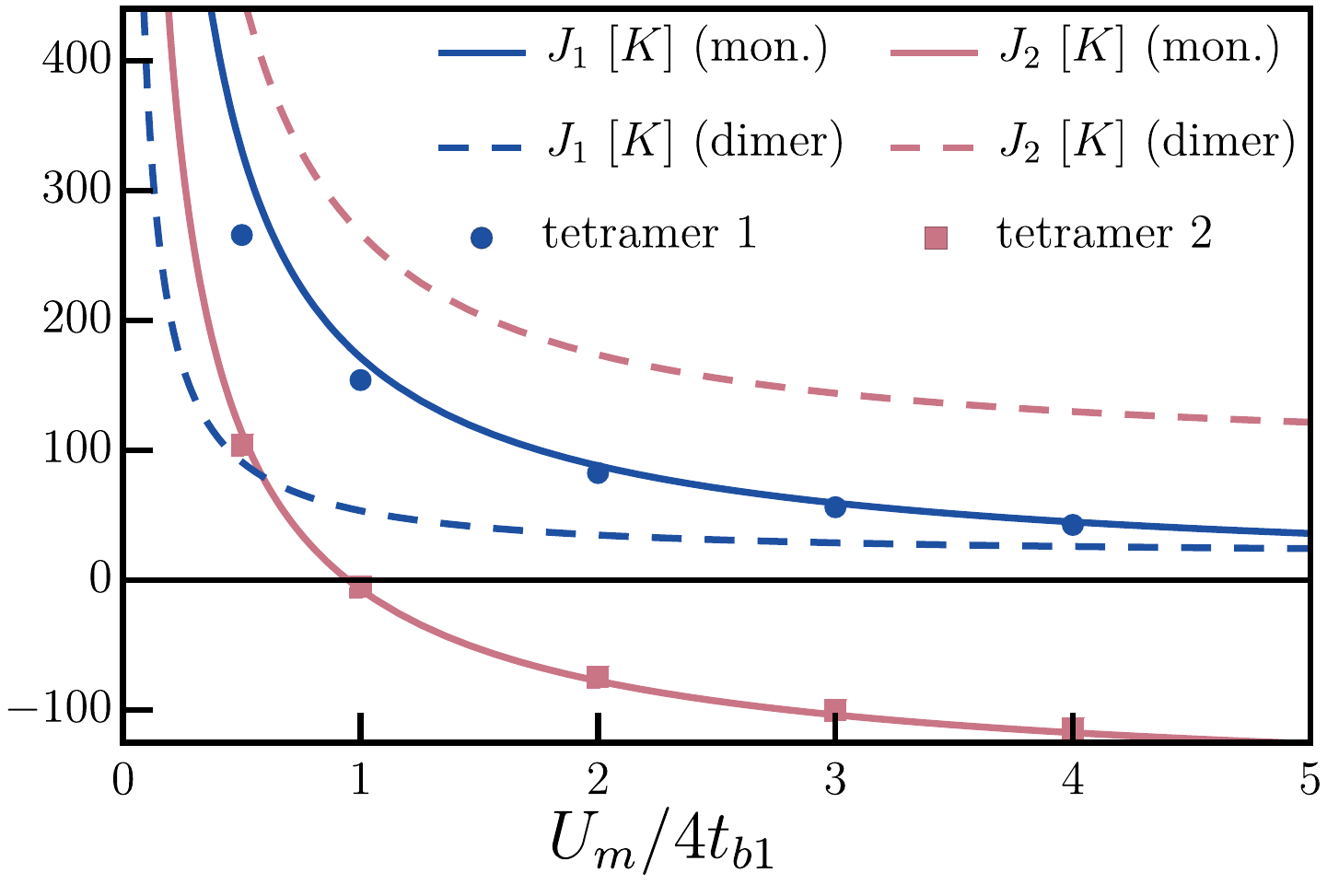}
	\end{center}
	\caption{Superexchange from perturbation theory for  the monomer (solid lines) and dimer (dashed lines) models compared with the exact singlet-triplet splitting of the tetramers marked in Fig. \ref{fig:models}a (dots and squares). Tight-binding parameters as calculated from first principles for \Cl and  $V_m=0$. 
%
} 
	\label{fig:Js}
\end{figure}

The  monomer model yields a  larger  $J_1$ than the dimer model, Fig. \ref{fig:Js}. This can be straightforwardly  understood. The dimer with two electrons admits several low-lying excited states that allow for additional superexchange pathways, these are omitted from the dimer model. The exact energy differences between the lowest energy singlets and triplets for tetramer 1 (Figs. \ref{fig:models}a, \ref{fig:Js}) are in excellent agreement with the perturbative treatment of the monomer model but are very different from the $J_1$ calculated from the dimer model.

$J_2$ is calculated from the analogous treatment of the perturbation ${\mathcal H}_2$. Here the predictions of the monomer model are  strikingly different from the dimer model. $J_2$ is very rapidly suppressed by $U_m$ in the monomer model, indeed $J_2$ becomes ferromagnetic ($<0$) for only moderate $U_m$ at $V_m=0$, Fig. \ref{fig:Js}. Again a comparison with the exact low-energy states of tetramer 2 (Figs. \ref{fig:models}a, \ref{fig:Js}) demonstrates excellent agreement with the monomer model and profound differences from the dimer model. 
In the monomer model, $J_2$ remains finite and negative whereas $J_1\rightarrow0$ as $U_m\rightarrow\infty$.

\begin{figure}
	\begin{center}
		\includegraphics[width=0.9\columnwidth]{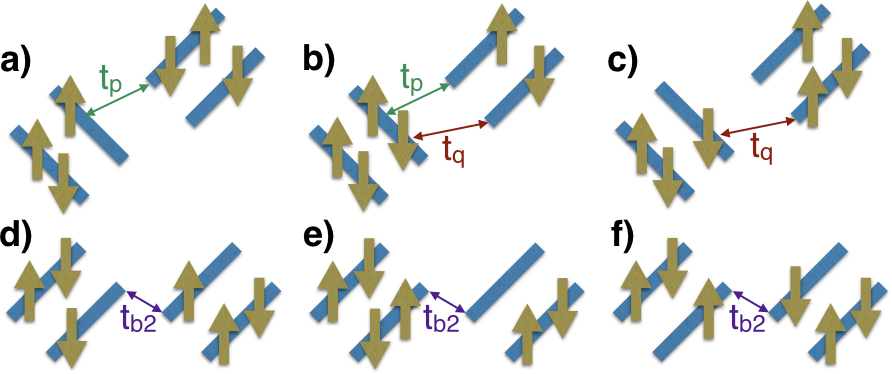}
	\end{center}
	\caption{Classical sketches of exchange process. 
		(a-c) An exchange pathway that remains finite as $(U_m-V_m)/ t_{b1}\rightarrow\infty$ contributing  $\propto t_pt_q/t_{b1}$ to $J_2$.
    (d-f) An exchange pathway that vanishes as $U_m\rightarrow\infty$, contributing  $\propto |t_{b2}|^2/U_m$ to $J_1$ for $U_m-V_m\gg t_{b1}$.
    } 
	\label{fig:process}
\end{figure}

\textit{Why is $J_2$ so different from $J_1$?} The essential difference is that there are two hopping pathways in $\mathcal{H}_2$ and only one in $\mathcal{H}_1$. This allows destructive interference between the different exchange pathways that contribute to $J_2$, which are necessarily absent in the calculation of $J_1$. Furthermore, processes with amplitudes $\propto t_pt_q$ can take place without incurring an energetic penalty $\propto U_m$, Fig. \ref{fig:process}. Thus, such processes remain active even as $U_m\rightarrow\infty$. Processes $\propto t_pt_q$ can favor ferromagnetic interactions.
To understand this, it is helpful to consider two limiting cases:  

(i) Molecular limit: $(U_m-V_m)/ t_{b1}\rightarrow\infty$.
A detailed understanding can be gained from considering the matrix elements
\begin{eqnarray}
M_{1} &=& 
\sum_{n} \frac{\bra{\uparrow_i\downarrow_j} \hat T_{21}^\dagger \ket{\Psi_n}\bra{\Psi_n} \hat T_{21} \ket{\downarrow_i\uparrow_j}}{2E_0(1)-\varepsilon_n} \notag \\
&=& \frac{1}{16t_{b1}} (\bra{S_i} - \bra{T_i}) (\ket{S_i} + \ket{T_i}) = 0, 
\\ 
M_{2} &=& 
\sum_{n} \frac{\bra{\uparrow_i\downarrow_j} \hat T_{21}^\dagger \ket{\Psi_n}\bra{\Psi_n} \hat T_{22} \ket{\downarrow_i\uparrow_j}}{2E_0(1)-\varepsilon_n} \notag\\
 &=& - \frac{1}{16t_{b1}}  (\bra{S_i} - \bra{T_i}) (\ket{S_i} - \ket{T_i}) = - \frac{1}{8t_{b1}},  \hspace*{0.8cm}
\end{eqnarray}
where 
$\ket{\sigma_i} = \frac{1}{\sqrt{2}}
\hat c_{i1\sigma}^\dagger 
\hat c_{i2\sigma}^\dagger 
(\hat c_{i1\overline{\sigma}}^\dagger
+\hat c_{i2\overline{\sigma}}^\dagger) \ket{0}$, 
$\ket{S_i} = \frac{1}{\sqrt{2}}
(\hat c_{i1\uparrow}^\dagger 
\hat c_{i2\downarrow}^\dagger -
\hat c_{i1\downarrow}^\dagger
\hat c_{i2\uparrow}^\dagger) \ket{0}$,  
$\ket{T_i} = \frac{1}{\sqrt{2}}
(\hat c_{i1\uparrow}^\dagger 
\hat c_{i2\downarrow}^\dagger  +
\hat c_{i1\downarrow}^\dagger
\hat c_{i2\uparrow}^\dagger) \ket{0}$
and $({\cal H}_{b1}^{(i)}+{\cal H}_{b1}^{(j)})\ket{\Psi_n}=\varepsilon_n\ket{\Psi_n}$.
$\ket{S_i}$ and $\ket{T_i}$ become degenerate as $(U_m-V_m)/ t_{b1}\rightarrow\infty$.
In the effective  Heisenberg model  $J_1=2t_{b1}^2M_{1} + \dots$ and $J_2=2t_pt_qM_2 + 2t_p^2M_1 + \dots$, where the  ellipses include  other terms at the same order, discussed  below.

$M_{1}$ vanishes because the intermediate singlet and triplet excited states interfere destructively, whereas  $M_{2}$ remains finite because the interference is constructive.
All other contributions to $J_1$ vanish due to similar interference effects, thus  $J_1=0$. In contrast, the dimer model predicts that $J_1\propto t_{b2}^2/t_{b1}$ in this limit.
All terms in $J_2$ proportional to $t_p^2$ and $t_q^2$ also vanish by the same arguments. Including all terms at this order yields $J_2=-t_pt_q/2t_{b1}$. 

(ii) \textit{In the $U_m=V_m$ limit}  the Hartree-Fock approximation becomes exact. This makes it straightforward to calculate the effective Heisenberg interaction, $J_\text{gen}$, for the more general perturbation $\mathcal{H}_\text{gen}=-\sum_{ij\mu\nu\sigma} (t_{\mu\nu}\hat T_{\mu\nu} + H.c.)$. One finds that $J_\text{gen}=2(t_{11}-t_{12}-t_{21}+t_{22})^2/{U_m}$.
Thus, $J_1=2t_{b1}^2/U_m$ and $J_2=2(t_p-t_q)^2/U_m$.
In this limit the interference  is a single particle phenomenon arising from the different phases of the two sites in the antibonding orbital.
Thus, the details of the interference  here are quite different from the molecular limit.
Nevertheless, one again finds that in the monomer model interference effects significantly suppress $J_2$ relative to expectations of the dimer model, where $J_2\propto(t_p+t_q)^2$.

\begin{figure}
	\begin{center}
		\includegraphics[width=0.8\columnwidth]{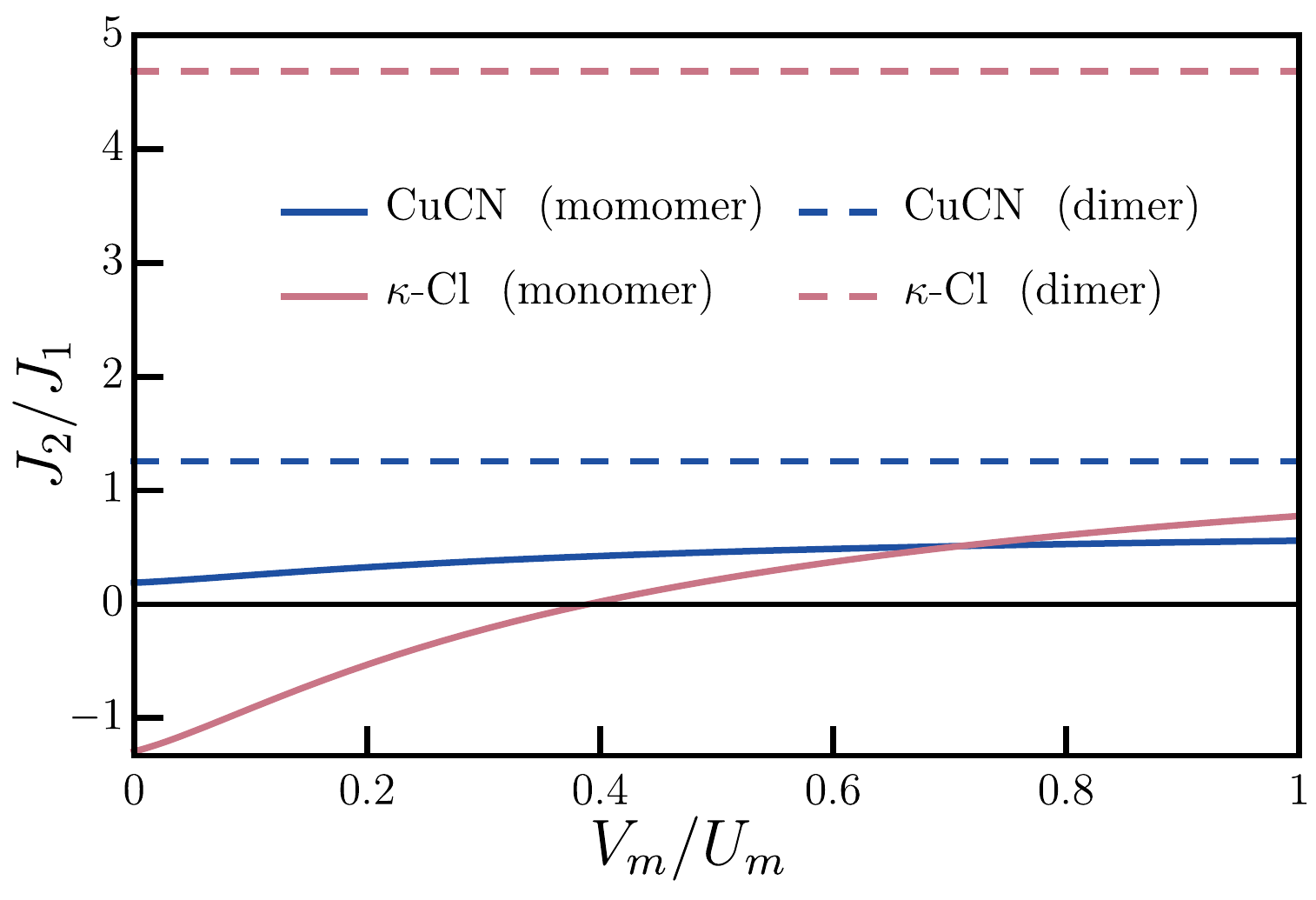}
	\end{center}
	\caption{Comparison of dimer (dashed lines) and monomer (solid lines) models for \CN  and \Cl (hopping integrals from \cite{Koretsune} and $U_m=12t_{b1}$). The dimer approximation predicts lattices between the square ($J_2/J_1\rightarrow\infty$) and triangular ($J_2/J_1=1$) limits, whereas the monomer model gives lattices in the quasi-1D regime ($|J_2/J_1 |<1$) for reasonable parameters (say, $1/3\lesssim V_m/U_m\lesssim2/3$). } 
	\label{fig:JV}
\end{figure}

 $J_2>0$  for all $U_m=V_m$. More generally, increasing $V_m$ suppresses  ferromagnetic exchange and eventually drives it antiferromagnetic, Fig. \ref{fig:JV}.  Large ferromagnetic $J_2$ is inconsistent with experiment. This suggests that $V_{m}/U_m$ is reasonably large, consistent with first principles estimates \cite{ScrivenPRB,Laura1}.

To consider specific materials we take the hopping integrals from previous first principles calculations \cite{Koretsune}. For reasonable parameters the monomer model yields $J_2<J_1$ in marked contrast to the dimer model, which gives $J_2>J_1$,  Fig. \ref{fig:JV}. We will describe this as dynamical dimensionality reduction (DDR). We will see below that DDR, and the frustration inherent in the system, leads  to a natural interpretation of the spin liquid phase in terms of coupled chains.
Another important difference from the predictions of the dimer model is that the values of the interactions, $U_m$ and $V_m$, are vital for determining the parameters of the Heisenberg model. These are \textit{not} well known at present and may differ between materials, but the best  estimates suggest that $U_m\gg t_{b1}$ and $V_m\lesssim U_m$ \cite{ScrivenPRB,Laura1,Laura2,Mori}.


The similarities between the hopping integrals in the BEDT-TTF and Pd(dmit)$_2$ salts suggest that similar physics is at play in the latter. Again the effective Heisenberg model is given by Eq. (\ref{Eq:Heis}) and there are two significant inter-dimer hopping pathways that contribute to $J_2$, but a single  pathway dominates $J_1$ 
 \cite{ScrivenPRL}.

{
In the 1D limit ($J_2\rightarrow0$) a paramagnetic Tomonaga-Luttinger liquid (TLL)   is expected at low temperatures. Eq. (\ref{Eq:Heis}) with $J_1>J_2>0$  received extensive attention \cite{LSWT,StarykhPRB,StarykhPRL,Sorella,Bursill,Singh,White,Bishop,Zheng99,Kohno,JOF06,JOF07,RVB,Bocquet} following the observation of a strong inelastic continuum, consistent with deconfined spinons, in neutron scattering experiments on Cs$_2$CuCl$_4$ \cite{Radu}, where $J_2\simeq0.34J_1$ \cite{Radu-model}.
Cs$_2$CuCl$_4$ displays spiral order at low temperatures. Nevertheless, the observed inelastic continuum  is quantitatively reproduced by theories based on disjunctive TLLs \cite{Kohno}. 
}

Classically, model (\ref{Eq:Heis}) has spiral order in the chain limit \cite{LSWT}. Quantum fluctuations enhance the one-dimensionality of this state \cite{LSWT,StarykhPRB,Singh,White,Zheng99,JOF06,JOF07,RVB}. Indeed Starykh \etal  argued that the model is q1D for $J_2<0.7J_1$ \cite{StarykhPRB}. Numerical studies are particularly challenging because of the incommensurate wavevector that characterizes the spiral phase  \cite{Singh,White} and several other ground states are  found to be energetically competitive \cite{StarykhPRL,Bishop,Zheng99,Sorella,Bursill,Singh,White}.

However, this question may be academic: theory suggests that small interactions  decide which competing phase is realized  \cite{StarykhPRB,JaimeSB}, as one expects on general grounds in frustrated systems. Series expansions  \cite{Singh} find that if the magnetization does not vanish as $J_2\rightarrow0$ then it becomes small extremely rapidly, consistent with 
the N\'eel temperature, $ T_N\sim\exp{[-(J_1/J_2)^2]} $, predicted from treating the intrachain dynamics via TLL theory  and the interchain coupling via the random phase approximation (TL+RPA) \cite{Bocquet}.

{
Therefore, our prediction that $J_1> J_2$ naturally explains the absence of long-range magnetic order in \CN and \Ag. Namely, that the q1D limit survives even for relatively large $J_2/J_1<1$ and thus the spin liquid is a remnant of the TLL found in an isolated chain. Even if the materials order eventually, the exponential suppression of $T_N$ can easily move this orders of magnitude below the lowest temperatures  studied (10s of mK).
\textit{Why then is  \Cl antiferromagnetic?} 
Two  perturbations are formally relevant \cite{StarykhPRB}: interlayer exchange, $J_z$; and the staggered interlayer component of the interchain \DM (DM) interaction, $D$, cf. Fig. \ref{fig:models}c \cite{Kagawa,SmithL,SmithB,Antal,Winter} 
(an inversion center precludes  DM coupling within the chains). 
}

\begin{figure}
	\begin{center}
		\includegraphics[width=0.9\columnwidth]{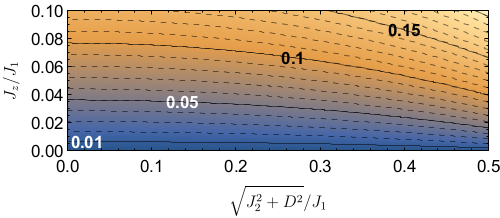}
	\end{center}
	\caption{Calculated N\'eel temperature, $T_N/J_1$, for the Heisenberg model with interchain coupling treated at the RPA level. For reasonable parameters, a critical temperature $\sim20$~K can be realized (cf. Fig. \ref{fig:Js}), as observed in \Cl.} 
	\label{fig:TN}
\end{figure}

In the TL+RPA theory \cite{Bocquet} the dynamic susceptibility is given by $\chi^{+-}_{3D}(\omega,{\bm k})=\chi_{1D}^{+-}(\omega,k_x)/[1-\tilde{J}(\bm k)\chi_{1D}^{+-}(\omega, k_x)]$, where $\chi_{1D}^{+-}(\omega, k_x)$ is the  susceptibility of a single chain perpendicular to $D$ and the Fourier transform \cite{foot-bi} of the interchain interactions is $\tilde{J}(\bm k)=-J_z\cos k_z\pm\sqrt{J_2^2+D^2}[\cos(k_y/2)+\cos(k_x-k_y/2)]$  . $T_N$ is the highest temperature with a zero-frequency pole in $\chi^{+-}_{3D}(\omega,{\bm k})$. This is straightforwardly calculated as described in \cite{Bocquet}. The solutions, 
%
 Fig. \ref{fig:TN}, clearly indicate that for reasonable parameters it  is possible to achieve $T_N/J_1\sim 0.1-0.2$, consistent with  observed critical temperature ($\sim20$~K) in \Cl, given our calculation of $J_1$, Fig. \ref{fig:Js}. Furthermore, $J_z$, which is unfrustrated, affects $T_N$ far more strongly than  $D$ or $J_2$,  suggesting this could be the essential difference between \Cl and \CN.
{ This could be tested by applying uniaxial strain perpendicular to the layers, which one would expect to increase $J_z$. This should increase $T_N$ in \Cl and perhaps even drive \CN or \Ag antiferromagnetic for sufficiently large strains, if the Mott transition does not intervene.
	Two intermonomer hopping integrals are relevant to interlayer hopping \cite{Jacko}, so the  interference effects that suppress $J_2$ also affect $J_z$. Thus, different materials may have radically different $J_z$.}


It is important to ask how the DDR picture of the dimer Mott insulating organics compares with experiment. 

{
{Heat capacity}   varies linearly with temperature in a TLL \cite{Giamarchi} as  observed in  \CN  \cite{Yam08}.  
Thermal conductivity of \CN does not reveal a term that varies linearly with temperature \cite{Yam09}.
It has widely been assumed, on the basis of q2D theories, that this is inconsistent with the heat capacity measurement.
However, in a weakly disordered spin chain the magnetic contribution to the thermal conductivity $\kappa_\textrm{mag}\propto T^2$ \cite{Rosch}, which is  consistent with the measurements of  \CN provided the magnetic contribution dominates the low temperature behavior \cite{Yam09}. One also expects a dip in $\kappa_\textrm{mag}$ at $g\mu_BB\sim 4k_BT$ \cite{Rosch}, which is also observed \cite{Yam09SI}.
}

At low-frequencies one expects  a power-law in the optical conductivity of a TLL \cite{Giamarchi}. This is  observed in both \CN \cite{Dressel12} and \Ag \cite{Dressel17}.

The bulk susceptibility of \CN shows a broad maximum around $\sim70$ K \cite{Shimizu}. This can be fit reasonably well by high temperature series expansions for the isotropic triangular lattice \cite{Shimizu,Zheng}. However, for 1D chains one also expects a broad maximum at $T=0.64 J$ \cite{Affleck}, which would lead to the estimate $J_1\sim100$~K in \CN.
%

%
The nuclear magnetic resonance (NMR) relaxation rate, $1/T_1$, in \Cl is well understood in terms of the incipient magnetic order \cite{Yusuf1,Yusuf2}. In contrast, $1/T_1$
in \CN is a long-standing problem  \cite{Lee,Qi08,Qi09,Baskaran,Liu,Kyung,Grover,Galitski,Kawamura,Li}. $1/T_1$ in \CN decreases as the temperature is lowered until a minimum is reached at $\sim 6$~K. A broad peak is then observed around 1~K. 
In spin chains one expects a  minimum in $1/T_1$ at $T\sim J_1/10$ concomitant with the crossover to a TLL  \cite{Dupont}.
There are no calculations, to date, describing $1/T_1$ in the presence of interchain interaction at temperatures above the TLL regime. 
Therefore, the only available comparison is with experimental results for Cs$_2$CuCl$_4$ \cite{Vachon}.
Note that $J_2/J_1$ and the  DM interaction are similar, but not identical, in the two materials, so the analogy is imperfect. However, one does not expect charge fluctuations to be especially important in the organics as no dramatic changes are observed under pressure until the first order metal-insulator transition.
In Cs$_2$CuCl$_4$ one observes a broad peak around $\sim2.5$~K, associated with the emergence of short-range order (SRO) \cite{Vachon,Tokiwa}, that is strongly reminiscent of the peak at $\sim 1$~K in \CN. {Microscopically, this SRO may be associated with the binding of spinons into triplons \cite{Kohno,Balents} driving a dimensional crossover and cutting off the logarithmic divergence in $1/T_1$ expected in a TLL.  }

Therefore a natural explanation of $1/T_1$ in \CN is that one sees a high temperature regime, a crossover to a TLL regime at $T\sim  6$ K and  the emergence of SRO{/triplons} at $T\sim1$~K. 
These crossovers could also be responsible for the anomalies observed at the same characteristic temperatures in many other experiments \cite{Yam08,Yam09,Manna,Poirier12,Poirier14}.
A clear prediction of this interpretation is that the emergence of SRO should lead to the broadening of NMR spectra as the temperature is lowered \cite{Vachon}. This is indeed observed in $^{13}$C NMR in \CN; an observation that has eluded explanation in q2D theories  \cite{Shimizu06}.

Antal \etal recently concluded that electron spin resonance (ESR) ``in \Cl resembles the ESR in 1D Heisenberg chains with a Dzyaloshinskii-Moriya interaction''  \cite{Antal}
just as our calculations  suggest.
Therefore, our prediction that the spin correlations in the insulating state are q1D is consistent with many experiments.

Metallic organics display coherent in-plane electronic transport  at low temperatures. DDR applies only to the spin correlations and so is not inconsistent with this. However, charge transport becomes incoherent above 20-40~K \cite{Broun,Jaime-Dressel}.
This suggests that the coherent interference processes, responsible for DDR, may be washed out when the temperature is raised. This would imply a strong temperature dependence in $J_2/J_1$ and  lead to a dimensional crossover at a much lower temperature scale than one would expect from the low temperature $J_2/J_1$.

$J_2$ favors $d_{x^2-y^2}$ superconductivity (taking the $x$ and $y$ axes to lie along the $J_2$ bonds), whereas $J_1$ favors $s+d_{xy}$ pairing  \cite{RVB,group}. Experimentally, the pairing symmetry in the organics remains controversial, but our results appear to favor $s+d_{xy}$ superconductivity, perhaps with accidental nodes.

Cs$_2$CuCl$_4$ displays a  rich phase diagram as the strength and orientation of the magnetic field is varied \cite{StarykhPRB,Tokiwa,Vachon}. Therefore, a more complete mapping of the physics of the
organics in terms of field strength and direction, particularly those with antiferromagnetic order,
and a detailed
comparison with q1D theory, including the full details of
the DM interaction, would provide a powerful test of the
ideas described above.
So could quantitative understanding of magnetic Raman scattering \cite{Nakamura14,Nakamura16,DrichkoPRB,Drichko}.


\begin{acknowledgments}
We thank Anthony Jacko, Amie Khosla, and Ross McKenzie for helpful conversations. This work was supported by the Australian Research Council through Grants No. FT130100161 and DP160100060. J.M. acknowledges financial support from (MAT2015-66128-R) MINECO/FEDER, UE. 
\end{acknowledgments}

\section{Appendix: Erratum}

In the above Letter we presented a general calculation of the superexchange interactions in dimer Mott insulators and compared these results to the title materials. We contrasted the  `monomer model', with one  orbital per molecule, to the so-called dimer approximation where the bonding and antibonding combinations of the two orbitals within each dimer are first constructed and the orbitals that are filled in the non-interacting limit are neglected. We showed that intradimer interference effects can lead to qualitative differences between the full monomer model and the dimer approximation. Most dramatically, intradimer interference  can cause a quasi-one-dimensional Heisenberg model to arise as an effective low-energy model of  a quasi-two-dimensional tight-binding model. We argued that this physics is relevant to the $\kappa$-(BEDT-TTF)$_2X$.

However, we have subsequently discovered an important error in our application of the general theory to the BEDT-TTF salts. 
We wrote the tight-binding part of the Hamiltonian in the form $-\sum_{ij\sigma}t_{ij}\hat c_{i\sigma}^\dagger \hat c_{j\sigma}$, where $\hat c_{i\sigma}^{(\dagger)}$ annihilates an electron with spin $\sigma$ in the $i$th Wannier orbital. We will continue to use this convention throughout this erratum. However, Koretsune and Hotta \cite{Koretsune} write such terms in the form $+\sum_{ij\sigma}t_{ij}\hat c_{i\sigma}^\dagger \hat c_{j\sigma}$. We failed to account for this sign difference when using their first principles parameters for the $\kappa$-(BEDT-TTF)$_2X$ salts. This has important consequences for these materials. As discussed below, the general mechanism described in our Letter whereby a one-dimensional superexchange interaction results from a two-dimensional tight-binding model remains correct. However, it does not appear to be relevant to the $\kappa$-(BEDT-TTF)$_2X$ salts. 

The value of $J_1$ is independent of the signs  of the hopping integrals, but $J_2$ is not (cf. Fig. 1). As emphasized in our Letter, interference effects can dominate the value of $J_2$, particularly when electron-electron interactions are large. In our Letter we analyzed the superexchange interactions analytically, in two limits. We begin this erratum by clarifying how the superexchange interactions are changed when the signs of the hopping integrals are reversed (the numbering below corresponds to that on pages 2 and 3). 

(i) In the molecular limit, $(U_m-V_m)/|t_{b1}|\rightarrow\infty$, the analytic forms we reported above are correct regardless of the signs of the hopping integrals. In this limit the only non-vanishing superexchange interaction is $J_2=-t_pt_q/2t_{b1}$. Changing the signs of all three hopping integrals takes $J_2\rightarrow-J_2$. This is consistent with the differences between Nagaoka ferromagnetism and Haerter-Shastry antiferromagnetism \cite{Powell,Haerter,Sposetti}. However,  the signs of the hopping integrals that Koretsune and Hotta found via density functional theory (DFT) imply that in the molecular limit the $\kappa$-phase salts are described by a antiferromagnetic Heisenberg model on a \textit{square lattice} as $J_1=0$ and $J_2>0$. Thus, in this frequently studied limit our  conclusion that there are important  differences between the superexchange interactions calculated from the monomer and dimer models remains valid (the latter yields a Heisenberg model on the anisotropic triangular lattice in this limit \cite{RossReview}).

(ii) For $U_m=V_m$ changing the sign of $t_{b1}$ swaps the bonding and antibonding orbitals relative to the definitions on page 1. That is, with $t_{b1}<0$  the bonding orbital is created by $\hat b_{i\sigma}^\dagger=(1/\sqrt{2})(\hat c_{i1\sigma}^\dagger - \hat c_{i2\sigma}^\dagger)$ and the antibonding orbital is created by $\hat a_{i\sigma}^\dagger=(1/\sqrt{2})(\hat c_{i1\sigma}^\dagger + \hat c_{i2\sigma}^\dagger)$. These signs propagate through and change our analytical results. The general superexchange interaction in this limit for $t_{b1}<0$ is $J_\textrm{gen}=2(t_{11}+t_{12}+t_{21}+t_{22})^2/U_m$, which is importantly different from the expression for $t_{b1}>0$ given on page 3. The expression for $J_1$ is unchanged, but for $t_{b1}<0$ we have $J_2=2(t_p+t_q)^2/U_m$, which  again differs by a sign   from the expression for $t_{b1}>0$, given in our Letter. An important consequence of these corrections is that the ratio $J_2/J_1$ from the dimer approximation is  correct when $U_m=V_m$: observe that the monomer and dimer lines in Fig. \ref{fig-new-4}  coincide at $V_m/U_m=1$ for both materials. 

\begin{figure}
	\begin{center}
		\includegraphics[width=0.9\columnwidth]{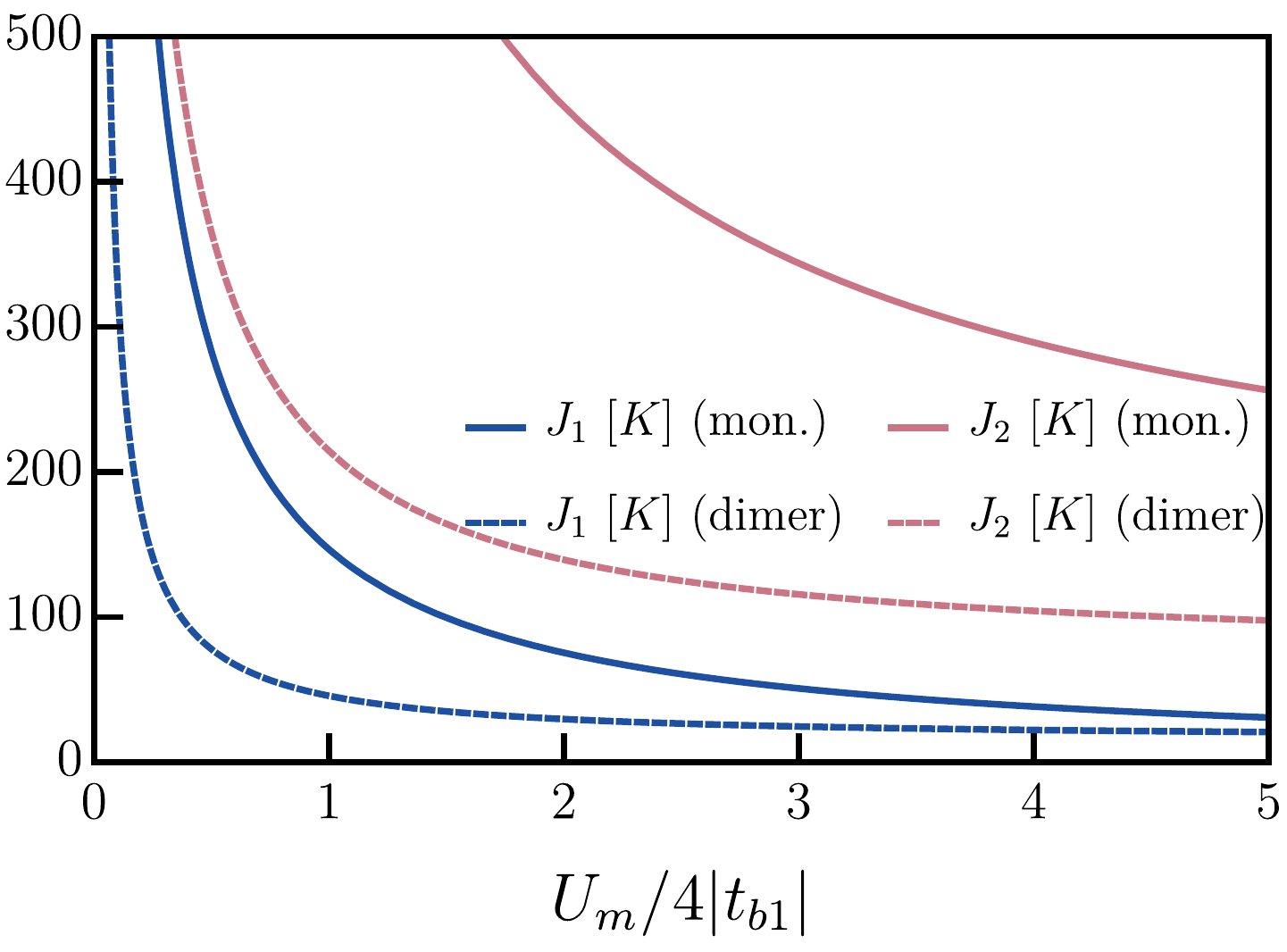}
	\end{center}
	\caption{Corrected version of Fig. 2: Values of the superexchange interactions for tight-binding parameters for \Cl \cite{Koretsune}. With our sign convention the hopping integrals are $t_{b1}=-207$~meV, $t_{b2}=-67$~meV, $t_p=102$~meV, and $t_q=43$~meV.}
	\label{fig-new-2}
\end{figure}

Our error also necessitates the correction of two figures from our Letter. 
In Figs. \ref{fig-new-2} and \ref{fig-new-4} we re-plot Figs. 2 and 4 with the signs of all the hopping integrals reversed (so as to correctly represent the results of DFT calculations \cite{Koretsune}). It can be seen from Fig. \ref{fig-new-2}  that both $J_1$ and $J_2$ remain antiferromagnetic ($>0$) for all values of $U_m$. This is in contrast to the case reported above, with the signs of all hopping integrals reversed, where $J_2$ becomes ferromagnetic ($<0$) for sufficiently large $U_m$. This behavior is expected for large $U_m$ as, to leading order in $1/U_m$, we have $J_2=-t_pt_q/t_{b1}$; changing the signs of all three hopping integrals must change the sign of $J_2$ in this limit. Fig. \ref{fig-new-4} shows that the superexchange interactions in both \CN and \Cl remain quasi-two-dimensional. Indeed, as one expects $U_m>V_m$, these results suggest that the dimer approximation \textit{underestimates} $J_2/J_1$ and hence that the magnetic interactions are closer to the square lattice than one would expect from the dimer approximation. This emphasizes that accurate estimates of the interaction parameters are important for determining the ratio $J_2/J_1$ and hence for understanding the spin liquid state in \CN.

Finally, to understand the role of the signs of hopping integrals in molecular Mott insulators more generally, the following observations may be helpful. Considering the pattern of hopping integrals (Fig. 1) given that the sign of $t_{ij}$ is reversed by a $\pi$ gauge transformation on exactly one of the sites ($i$ or $j$), changing only the sign of $t_{b1}$ is sufficient to change the results between those shown here and those reported above. Secondly,
as changing the signs of all hopping integrals is equivalent to a particle-hole transformation, if there is one electron per dimer (rather than three) the formulae given in the original Letter hold. This is the relevant filling for organic anion systems such as the Ni(dmit)$_2$ salts \cite{Kato}, which do appear to show quasi-one-dimensional magnetism.

Thus we conclude the following: (i) The interference mechanism for realizing a quasi-one-dimensional (q1d) Heisenberg model from a quasi-two-dimensional (q2d) tight-binding model is not relevant to the $\kappa$-phase organics, but is possible and may be realized in other materials. (ii) Inter-dimer interference effects are still likely to be relevant to the $\kappa$-phase organics, but they tend to drive the system towards the square lattice limit ($J_1/J_2\rightarrow0$). (iii) The dimer approximation is only accurate when the Hartree-Fock approximation is reasonable.

\begin{figure}
	\begin{center}
		\includegraphics[width=0.9\columnwidth]{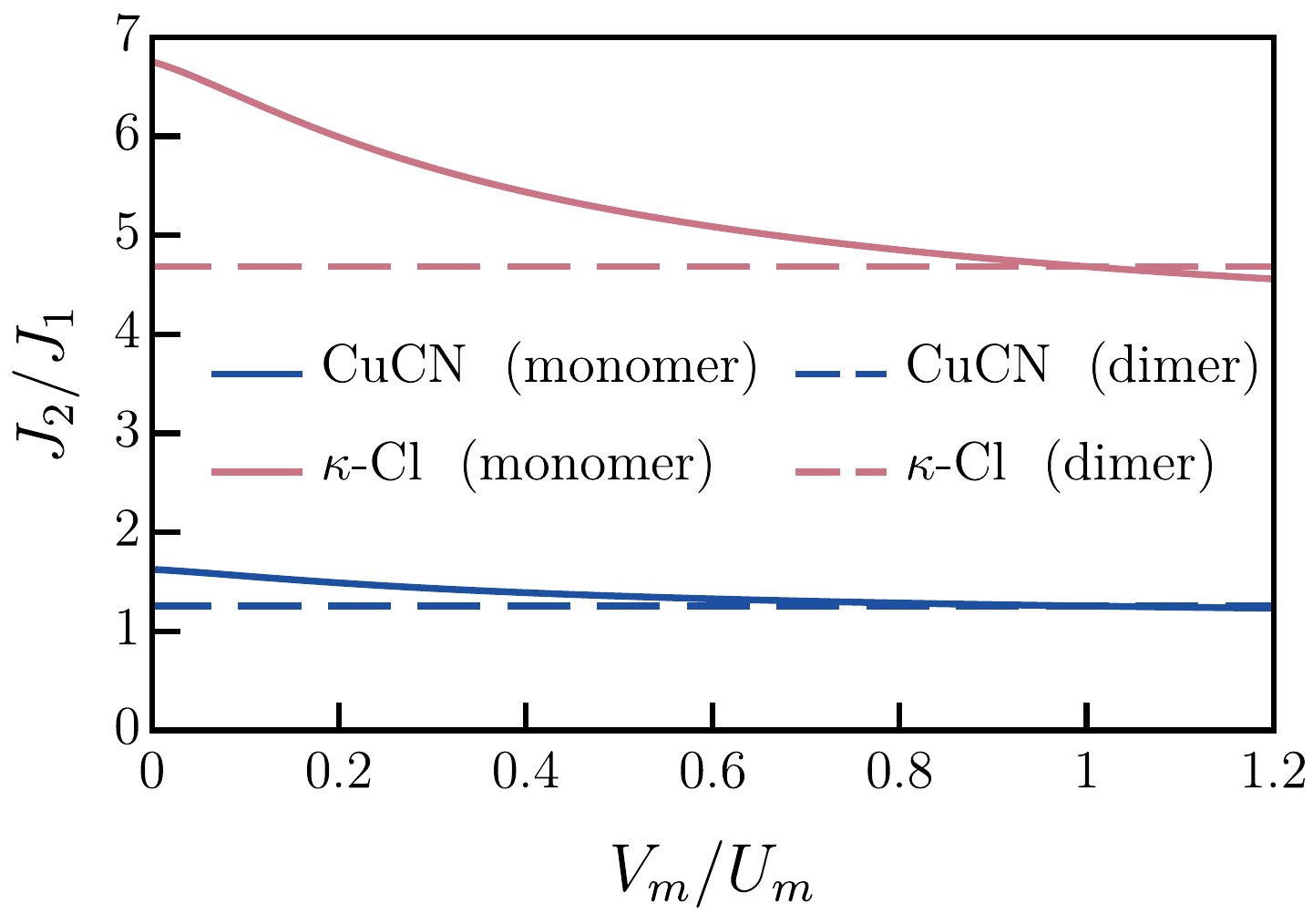}
	\end{center}
	\caption{Corrected version of Fig. 4. Ratios of the superexchange interactions with tight-binding parameters from \cite{Koretsune}. With our sign convention these are $t_{b1}=-207$~meV, $t_{b2}=-67$~meV, $t_p=102$~meV, and $t_q=43$~meV  for \Cl and $t_{b1}=-199$~meV, $t_{b2}=-91$~meV, $t_p=85$~meV, and $t_q=17$~meV  for \CN.}
	\label{fig-new-4}
\end{figure}

We thank Stephen Winter and Roser Valent\'i for helpful discussions.


\begin{thebibliography}{100}
\bibitem{foot-bi} 
We report our results in the Brillouin zone relevant before the unit cell is doubled by the pattern of DM vectors and  neglect the bilayer structure of \CN and \Ag as first principles calculations \cite{Winter} suggest that the DM vector is not staggered in this direction.
\end{thebibliography}
\end{document}